\documentclass[twocolumn,showpacs,amsmath,amssymb,superscriptaddress,aps,prc,nofootinbib]{revtex4-1}
  \usepackage{graphicx}
  \usepackage{epstopdf}
  \usepackage{epsfig}
  \usepackage{bm}
  \usepackage{color}
  \usepackage{xcolor}
  \begin{document}
  %\title{Halo size and separation energy in $^{22}$C from core recoil momentum distribution}
  %\title{Constraining halo size and neutron separation energy in $^{22}$C from core recoil momentum distribution}
  %\title{Constraining $^{22}$C halo with core recoil momentum distribution}
  \title{ Four-body eikonal approach to  three-body halo nuclei scattering}
  \author{L. A. Souza}
  \affiliation{Universidade Federal de Lavras, C. P. 3037, 37200-000, Lavras, MG, Brazil}
    \affiliation{Instituto de F\'\i sica, Universidade de S\~ao Paulo,
  C. P. 66318, 05314-970, S\~ao Paulo, SP, Brazil}
  \author{E. V. Chimanski}
  \affiliation{Instituto Tecnol\'ogico de Aeron\'autica, DCTA,
  12228-900, S. Jos\'e dos Campos, SP,~Brazil.}
  \author{T. Frederico}
  \affiliation{Instituto Tecnol\'ogico de Aeron\'autica, DCTA,
  12228-900, S. Jos\'e dos Campos, SP,~Brazil.}
  \author{B. V. Carlson} 
  \affiliation{Instituto Tecnol\'ogico de Aeron\'autica, DCTA,
  12228-900, S. Jos\'e dos Campos, SP,~Brazil.}
  \author{M. S. Hussein}
  \affiliation{Instituto de F\'\i sica, Universidade de S\~ao Paulo,
  C. P. 66318, 05314-970, S\~ao Paulo, SP, Brazil}
  \affiliation{Instituto Tecnol\'ogico de Aeron\'autica, DCTA,
  12228-900, S. Jos\'e dos Campos, SP,~Brazil.}
  \affiliation{Instituto de Estudos Avan\c cados, Universidade de S\~ao Paulo C. P. 72012, 05508-970 S\~ao Paulo, SP, Brazil}
  \begin{abstract}
  Halo nuclei projectiles can undergo into a fragmentation process when scattered by a nucleus target. The corresponding core from exotic nuclei 
  is usually  observed while the others fragments of the reaction are not. We use the recently proposed theory by \cite{Carlson17} 
  with a four-body  {(4B)} description of the inclusive breakup 
reaction where the 
projectile is $^{20}$C described like a two-neutron halo nucleus.  
  The  momentum of both neutrons are integrated out giving a generic description of the core angular distribution. 
  In this preliminary study we perform an analysis of the inclusive inelastic breakup cross-section using the eikonal approximation 
  for the distorted wave function of the projectile. A study of the inclusive inelastic cross-sections of $^{20}$C  from the collision with different targets are presented.  
  \end{abstract}
  %\vspace{-0.5cm}
  %\begin{keyword}
  %halo-nuclei, few-body physics, breakup reactions
  %\end{keyword}
  \maketitle

  \section{Introduction}
  %PACS {27.30.+t,21.10.Gv, 21.45.+v,11.80.Jy, 21.10.Dr}
  %27.30.+t Properties of specific nuclei listed by mass ranges  20 = A = 38
  %21.10.Gv Nucleon distributions and halo features
  %11.80.Jy Many-body scattering and Faddeev equation
  %21.45.+v Few-body systems
  %21.10.Dr Binding energies and masses

  %%%%%% 
  %
  The neutron halo of unstable nuclei close to the drip line 
  has been observed inderectly by some experimental groups 
  \cite{Togano16, Pesudo17} using different reaction processes. 
  Some works like those investigate Borromean two-neutron halo nuclei as, for 
  example $^{22}$C, and non-Borromean type such as $^{20}$C, which are
  interesting by the fact that the many-body effects can be disentangled 
  from the few-body dynamics. The large spatial distribution of the 
  weakly bound halo of these exotic nuclei should bring some distinctive 
  features to the reaction cross-sections, and
 the observed large reaction cross-sections were associated to a possible extended structure of two-neutron halo in the carbon isotopes.%

  To contribute to the investigation of the two-neutron halo nuclei collision
  problem, we address the theoretical analysis of inelastic breakup reactions 
  when weakly bound nuclei are employed as projectiles.
 The four-body theory proposed to describe this kind of reactions, namely  
 the formalism developed to calculate inclusive breakup cross-sections and 
 used here was described in details in Ref.~\cite{Carlson17}. 
  This new theory can be applied to treat reactions with stable/unstable 
  projectiles composed by three-fragments {  (3B)}, such as 
two-neutron halo 
nuclei.
  It is an extension of the  inclusive three-body breakup model 
  for incomplete fusion 
  reactions developed for two fragment projectiles by Ichimura, Autern and 
  Vincent (IAV) \cite{IAV1985}, Udagawa and Tamura (UT) \cite{UT1981} and
  Hussein and McVoy (HM) \cite{HM1985}. 

  The relevance of the four-body theory developed in \cite{Carlson17} comes 
  from the fact that it allows to obtain the fragment (neutron) yield 
  in the reaction. In this framework, the inclusive breakup cross-section 
  is a sum of the four-body elastic plus  inclusive inelastic breakup 
  cross-sections involving the absorption cross-sections of the neutrons.

  %and which generalizes the three-body formula reviewed in Austern, et al. \cite{Aust}.
  %The connection between IAV, UT and HM theories was shown in \cite{Huss90}.

  Particularly interesting is the contribution of the two-fragment 
  correlation to the  inclusive inelastic breakup cross-section 
  through a three-body absorptive interaction, which appears naturally in the 
four-body formulation. 
  We present preliminary results by showing an eikonal analysis with the 
  S\~ao Paulo potential used to describe the core-target interaction 
  in $^{20}$C reactions with a target. 
  The internal wave function of the incoming projectile  is obtained 
  from a three-body renormalized model \cite{AdhPRL95}. 
  We compute the inclusive inelastic breakup cross-section in 
  which the halo neutrons, labelled as $x_1$ and $x_2$, are not observed. 
  The cross-sections are estimated for the collisions of a $^{20}$C 
  projectile with $^{12}$C  and  $^{208}$Pb targets $(A)$, 
  schematically represented as
  $A(a,b)X$, where $a = x_1 + x_2 + b$ and $b$ is the core. 
%  The Coulomb interaction contribution is calculated as well.

  %\textcolor{blue}{
  %The Halo nuclei wave function is obtained (lucas paper) an used in what follows. This, play an important role in the source function present in the Husssein etc description.
  % 
  %We follow a recently developed theory to treat the inclusive breakup of three-fragment weakly bound nuclei \cite{Carlson17}. 
  %The natural three fragment candidates for projectiles are Borromean, two-nucleon and unstable three-fragment halo nuclei. 
  %
  %The new formula contains the four-body dynamics both in the elastic breakup cross-section and in the inclusive inelastic breakup ones.
  %It can be applied to treat reactions with stable/unstable projectiles composed of three-fragments, for instance, 
  %weakly bound Borromean and two-nucleon halo nuclei. For these systems the fingerprints of Efimov physics\cite{Efim} and universality \cite{TF} could be 
  %revealed through the appearance of long-range correlations between the three-fragments. In addition, one can seek for the generalization of the surrogate method 
  %applied to reactions like $(d,p)$ and $(d,n)$ to the reactions $(t,d)$ and $(^3$He$,d)$, among others.
  %We studied the single fragment inclusive cross-section part of inelastic breakup $^{20}$C$+ $C$ =$$^{18}$C$+ X$ processes. 
  %}

  %pbox{0.9\textwidth}{5mm}{linewidth=2mm,framearc=0.1,linecolor=lightblue,fillstyle=gradient,gradangle=0,gradbegin=white,gradend=whiteblue,gradmidpoint=1.0,framesep=1em}
  \section{Theoretical Formalism}
We provide a briefly description of the formalism used, 
  for a complete discussion see \cite{Carlson17}.  

  The inclusive breakup cross-section is a sum of two distinct terms, the
  elastic breakup and the inelastic breakup cross-sections
  \begin{equation}
  \frac{d^{2}\sigma_b}{dE_{b}d\Omega_{b}} = \frac{d^{2}\sigma^{EB}_b}{dE_{b}d\Omega_{b}} + \frac{d^{2}\sigma^{INEB}_b}{dE_{b}d\Omega_{b}} \nonumber
  \end{equation}
  our interest is second quantity, the inelastic part of the reaction spectra. This can be computed by
  \begin{equation}
  \frac{d^{2}\sigma^{INEB}_b}{dE_{b}d\Omega_{b}} = 
  \frac{2}{\hbar v_a}\rho_{b}(E_b) \langle \hat{\rho}_{{{x}_1}, {{x}_2}}|{W_{x_1} + W_{x_2}} + { W_{3B}}|\hat{\rho}_{{{x}_1}, {{x}_2}}\rangle  \label{s4inel},
  \end{equation}
  where $W_{x_1}$, $W_{x_2}$ and $W_{3B}$, are the imaginary part of the optical interaction potential. The position of the weakly bound neutrons of the incident nucleus provide the source function
  \begin{eqnarray}
  \hat{\rho}_{X}(\textbf{r}_{x_{1}},\textbf{r}_{x_{2}}) &=& (\chi_{b}^{(-)}|\Psi_{0}^{4B(+)}\rangle \\
  &=& \quad \int d\textbf{r}_{b}\left[\chi_{b}^{(-)}(\textbf{r}_{b})\right]^{\dagger}\Psi_{0}^{4B(+)}(\textbf{r}_{b}, \textbf{r}_{x_{1}}, \textbf{r}_{x_{2}})\nonumber
  \end{eqnarray}
  containing the four body (two neutrons + halo + target) description.
  The full four-body scattering state is 
  $
  \Psi_{0}^{4B(+)}(\textbf{r}_{b}, \textbf{r}_{x_{1}}  \textbf{r}_{x_{2}}),
  $ 
  which should be ideally obtained by solving the Faddeev-Yakubovsky 
  equations in the continuum 
  with optical potentials for the fragments and the target, 
  in addition to the inter-fragment potentials. 
  The boundary condition for the projectile-target four-body 
  scattering contains the
  initial three-body wave function of the halo projectile,
  which in particular can be obtained by the three-body renormalized model \cite{AdhPRL95}.

  The inelastic part of the breakup cross-section of the weakly bound projectile is given by
  \begin{eqnarray}
    \frac{d^{2}\sigma^{INEB}_b}{dE_{b}d\Omega_{b}} &=& \rho_{b}(E_b) \frac{k_a}{E_a}\bigg [ \frac{{E_{x_1}}}{{k_{x_1}}}{ \sigma_{R}^{x_1}} + \frac{{E_{x_2}}}{{k_{x_2}}}{\sigma_{R}^{x_2}} \nonumber \\
      &&\qquad + \frac{E_{CM}({{x}_1},{{ x}_2})}{(k_{{x}_1}+ k_{{x}_2})} { \sigma_{R}^{3B}}\bigg ],
  \label{ineb}
  \end{eqnarray}
where $a$ and $x$ correspond to projectile and fragment labels%\footnote{Weakly bound projectile : $E_{{x_i},Lab} = E_{a, Lab}(M_{{x_{i}}}/M_a)$ with $M_{a}$ and $M_{{x_i}}$.}
, respectively.
  Finally, the single fragment  inclusive cross-sections are obtained by
  \begin{small}
  \begin{equation}
  \sigma_{R}^{x_i} = \frac{{k_{x_i}}}{E_{{x_i}}} \langle \hat{\rho}_{{x}_1, {x}_2}|W_{x_i}|\hat{\rho}_{{x}_1, {x}_2}\rangle,
  \end{equation}
  \end{small}
  and the double fragment inclusive cross-section is
  \begin{equation}
  \sigma_{R}^{3B} = \frac{(k_{{x}_1}+ k_{{x}_2})}{E_{CM}({{x}_1},{{ x}_2})}\ \langle \hat{\rho}_{{x}_1, {x}_2}|W_{3B}|\hat{\rho}_{{x}_1, {x}_2}\rangle\, \label{sig3b}
  \end{equation}
  which represents the two-fragment irreducible inelastic processes.

  \section{Cross Sections}
  The source function computed with the  Hussein and McVoy (HM) model \cite{HM1985} is written as
  \begin{equation}
  \langle \textbf{r}_{{x}_1}, \textbf{r}_{{x}_2}|\hat{\rho}^{4B}_{HM}\rangle = \hat{S}_{b}(\textbf{r}_{{x}_1}, \textbf{r}_{{x}_2})\chi^{(+)}_{{x}_1}(\textbf{r}_{{x}_1})\chi^{(+)}_{{x}_2}(\textbf{r}_{{x}_2})
  \end{equation}
  where the four-body scattering state is approximated by the product of the distorted waves of the three
  fragments and the projectile bound state wave function. The $\hat{S}$-matrix can be given by (for simplicity, we drop the $x$ index for fragments)
  \begin{equation}\label{s}
  \hat{S}(\bm{r}_{x_{1}},\bm{r}_{x_{2}})=\int d\bm{r}_{b}\phi_{3B} (\bm{r}_{x_{1}},\bm{r}_{x_{2}},\bm{r}_b)
  e^{i\bm{q}\cdot \bm{r}_{b}}\langle\tilde{\psi}^{(-)}_{\bm{p}_{b'}}|{\psi}^{(+)}_{\bm{p}_{b}}\rangle (\bm{r}_b),
  \end{equation}
  where $\phi_{3B}$ is the incoming three-body wave function computed with the renormalized three-body model, projected in four-body coordinates now.
  The core distorted waves are computed with the straight-line trajectory in the eikonal approximation,
  \begin{multline}\label{eik}
  \hat S^{Eik}_{\textbf{p}_{b}^{\prime}, \textbf{p}_{b}}=
  \tilde{\psi}_{\bm{p}_{b}^{\prime}}^{(-)}\left(\bm{r}_{b}\right)\psi_{\bm{p}_{b}}^{(+)}\left(\bm{r}_{b}\right)\\=  
  \exp\Bigg[-\frac{i\mu}{\hbar p_{b}^{\prime}}\int_{-\infty}^{z_{b}}dz V\left(\sqrt{b_{b}^{2}+z^{2}}\right)\\
  -\frac{i\mu}{\hbar p_{b}}\int_{-\infty}^{z_{b}}dz V\left(\sqrt{b_{b}^{2}+z^{2}}\right)\Bigg].
  \end{multline}
The scattering dynamics for this case is shown in  Fig.~1, where cylindrical coordinates are employed with $r_b=\sqrt{z_b^2+b_b^2}$.
  \begin{figure}
  \includegraphics[scale=0.15]{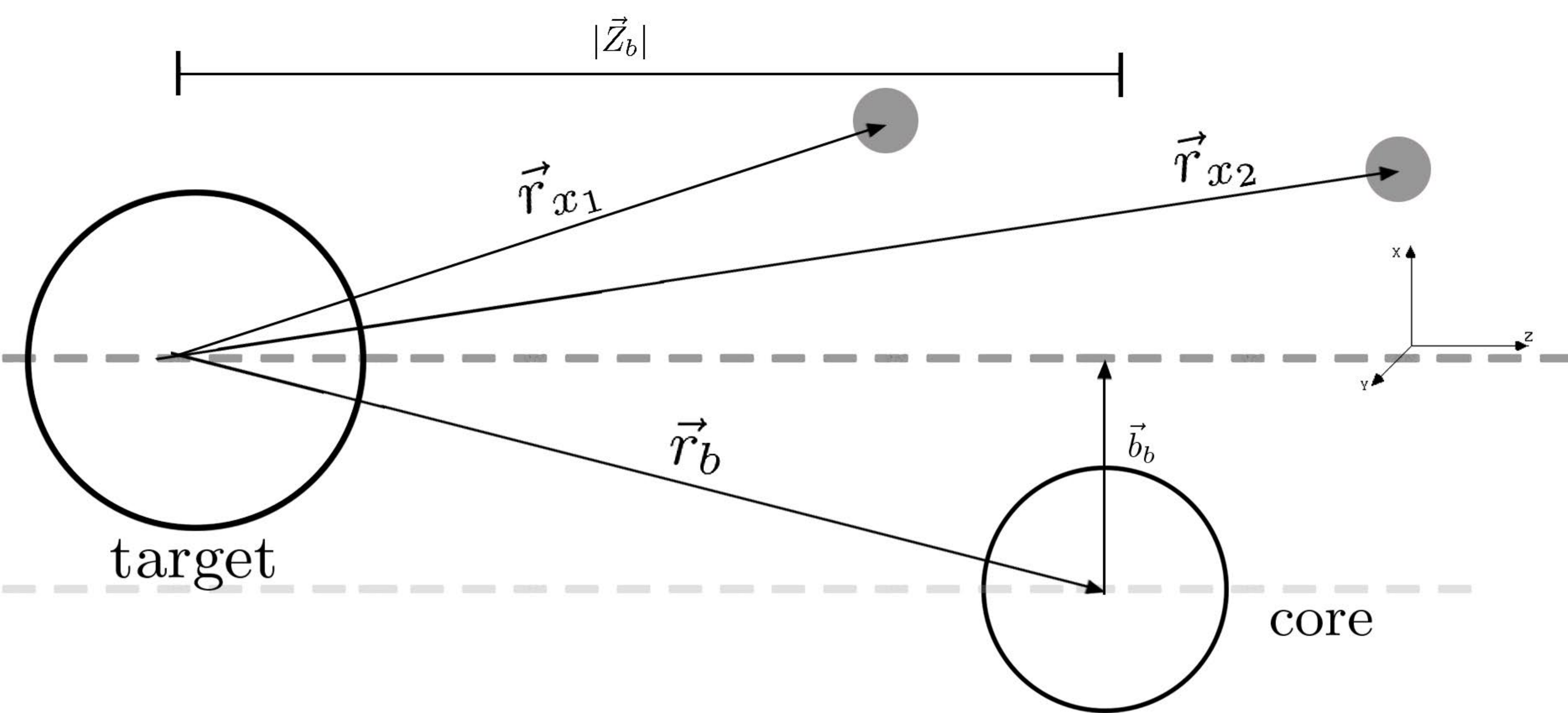}
  \caption{Diagram for the eikonal scattering approximation, the straight-line trajectory is along the z-direction in a cylindrical coordinate system.}
  \end{figure}
{  
The S\~ao Paulo (SP) optical potential is taken to represent the core-target
interaction $V$ in the Eq.~(\ref{eik}). Although the comparison to different 
optical potentials is out of the scope of this preliminary work, we stress that 
the heavy-ion S\~ao Paulo potential has been successfully applied to several 
cases, e.g. elastic,
inelastic and transfer reactions \cite{Chamon,Gasques}. The SP potential takes 
into account nucleon-nucleon interactions folded into nucleon densities 
products providing at the end a complex and energy depend interaction for a 
chosen nucleus.}
One of our goals is the study of the $\hat T$-matrix that can be obtained in 
the limit 
   \begin{equation}\label{teik}
  \hat T^{Eik}=\lim _{z_{b} \rightarrow \infty} \hat S^{Eik}-1.  
   \end{equation}
  For this, we let scattering to be in the $x-z$ plane and write
  \begin{eqnarray}
  \bm{q}_{b}\cdot\bm{r}_b&=&\bm{q}_{b}\cdot\bm{b}_b+\bm{q}_{b}\cdot z_b{\bm{ \hat k}}\nonumber\\
  &=& b_b\ k^\prime_b\sin\theta_b\cos\varphi_b+(k_b-k^\prime_b\cos\theta_b)z_b.
  \end{eqnarray}
   The single fragment inclusive cross-section in the four-body theory can 
then be obtained by
  \begin{small}
  \begin{multline}
  \frac{E_{{x}_1}}{k_{{x}_1}}\sigma_{R}^{x_1} = \int d\textbf{r}_{{x}_1} \int d\textbf{r}_{{x}_2} |\hat{S}_{b}(\textbf{r}_{{x}_1}, \textbf{r}_{{x}_2})|^{2}  |\chi^{(+)}_{{x}_2}(\textbf{r}_{x_2})|^2\\
  \, \times W^{nT}(\textbf{r}_{{x}_1})|\chi^{(+)}_{{x}_1}(\textbf{r}_{x_1})|^2 \, . \label{xx}
  \end{multline}
  \end{small}
   where $\kappa=1,2$ refers to the neutrons.
  % \newline
  % \begin{eqnarray*}\label{s2}
  %|\hat{S}(\bm{r}_{x_{1}},\bm{r}_{x_{2}})|^{2}&=&\left ( \int d\bm{r}_{b}\phi_{4B} (\bm{r}_{x_{1}},\bm{r}_{x_{2}},\bm{r}_b)
  %e^{i\bm{q}\cdot \bm{r}_{b}}\langle\tilde{\psi}^{(-)}_{\bm{p}_{b'}}|{\psi}^{(+)}_{\bm{p}_{b}}\rangle (\bm{r}_b) \right )^{\dagger}\\
  %&\times &\int d\bm{r '}_{b}\phi_{4B} %(\bm{r}_{x_{1}},\bm{r}_{x_{2}},\bm{r'}_b)%
  %e^{i\bm{q}\cdot %\bm{r'}_{b}}\langle\tilde{\psi}^{(-)}_{\bm{p}_{b'}}|{\psi}^{(+)}_{\bm{p}_{b}}\rangle (\bm{r'}_b)
  %\end{eqnarray*}
   %\newline
 
 {  
 The weakly bound nucleons of the projectile interact with the target 
nucleus via optical potential represented in this model by $W^{nT}$. Since the 
S\~ao Paulo potential was developed for nucleus-target interactions (meanly 
heavy nucleus), we represent, as a first approach, the neutrons-target 
interaction as a Woods-Saxon-like function. This, provides a good approximation 
for the interaction shape of potentials usually obtained by more sophisticated 
optical potentials. We take the Woods-Saxon potential as
}
  $$W^{nT}(r)=\frac{\ W_{0_I}}{1+\exp \left ((r-R)/a_I\right )},$$
  and the following parameters for interaction $n-^{12}$C: $W_{0_R}=49.9395$ MeV, $W_{0_I}=1.8256$ MeV, $a_R=a_I=0.676$ fm and $R_I=R_R=2.5798$ fm; from Ref.~\cite{Koning}. 
For the two-fragment irreducible inclusive cross-section, we have that
  \begin{small}
  \begin{multline}
  \sigma^{3B}_{R}= \int d\bm{r}_{x_{1}}\int d\bm{r}_{x_{2}} \left |\hat{S}(\bm{r}_{x_{1}},\bm{r}_{x_{2}}) \right |^{2}
   \left | \chi ^{(+)}_{x_{2}}(\bm{r}_{x_{2}})\right |^{2} \\
  \times W^{3B}(\bm{r}_{x_{1}},\bm{r}_{x_{2}})\left | \chi ^{(+)}_{x_{1}}(\bm{r}_{x_{1}})\right |^{2},
  \end{multline}
  \end{small}
  where Woods-Saxon potential is also used for $W^{3B}$ interaction.
  \section{Eikonal Analysis}
  The transition matrix obtained within the eikonal approximation (\ref{eik}) is presented in Figure 2. We show the $\hat T$-matrix computed for a $^{20}$C projectile 
  on the target of $^{12}$C as a function of the impact parameter $b_b$. The most of the absorption occurs around the target nucleus radius, vanishing for larger values of impact parameter (no Coulomb interaction was taken into account in this case). 
  \begin{figure}
  \includegraphics[width=0.85\columnwidth]{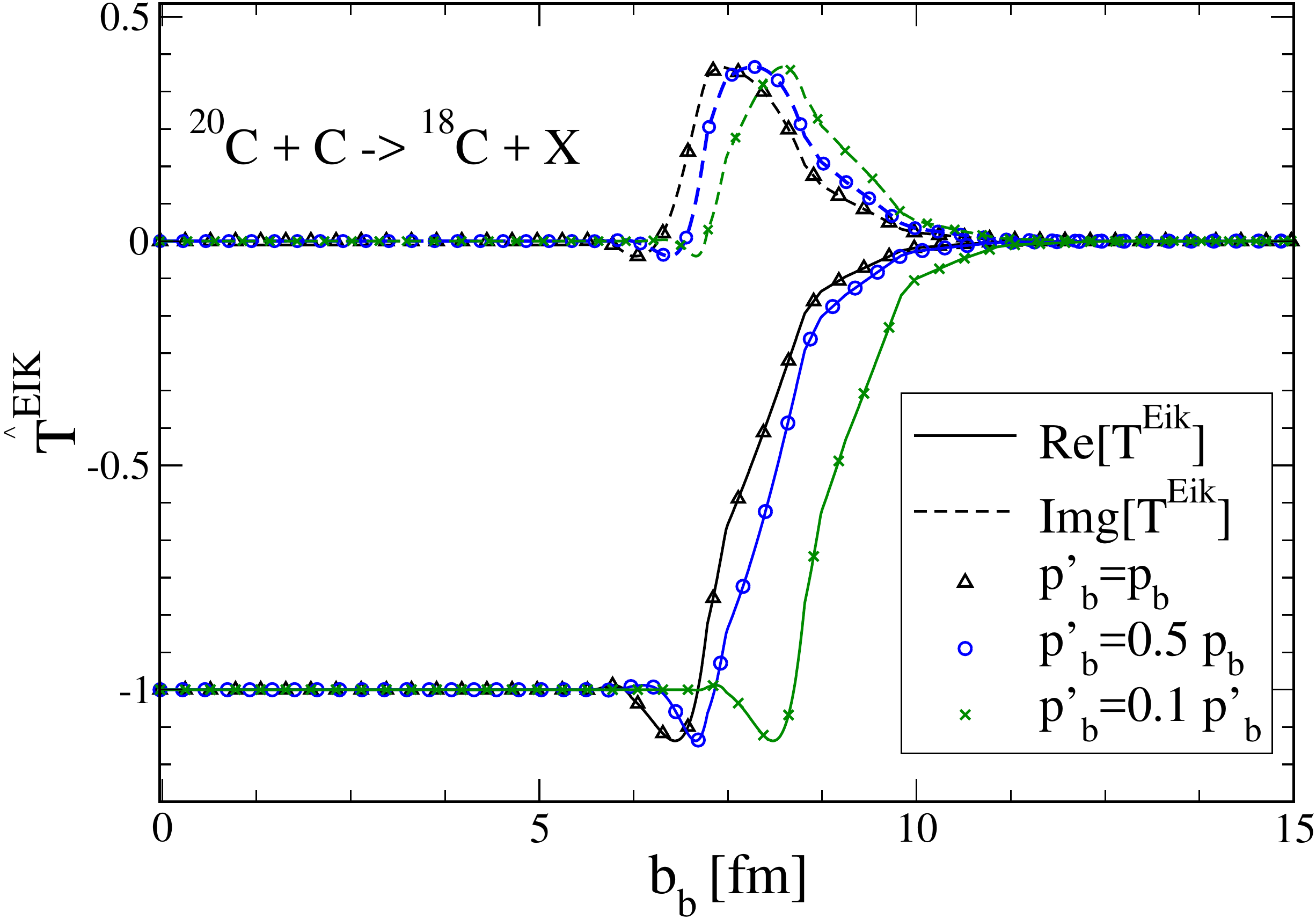}
  \caption{Real and imaginary parts of the scattering $\hat T$-matrix  for $^{20}$C off
  $^{12}$C at 50 MeV/nucleon, as a function of the impact parameter $b_b$. }
  \end{figure}
  We also analyzed the role of the three body wave function when coupled to the $\hat T$-matrix (we left out the transferred momentum plane wave). This term is present in the $\hat{S}({\bf r}_{x_{1}},{\bf r}_{x_{2}})$ amplitude. Fig.~3 presents this quantity for the two neutrons fixed with $b_1=b_2=8$ fm (both near the target radius). For bigger values of the impact parameter the contribution of this configuration vanish because the three body wave function is close to zero, since the two neutrons in the halo are far away from the core. As the core approaches the halo and this case also the target, the reaction takes place. This strengthen the importance of the three body wave function in the process. We note that all fragments positions have to be considered at the end, i.e. their respective volume integration have to be performed for the cross-section calculation. 
  \begin{figure}
  \includegraphics[width=0.85\columnwidth]{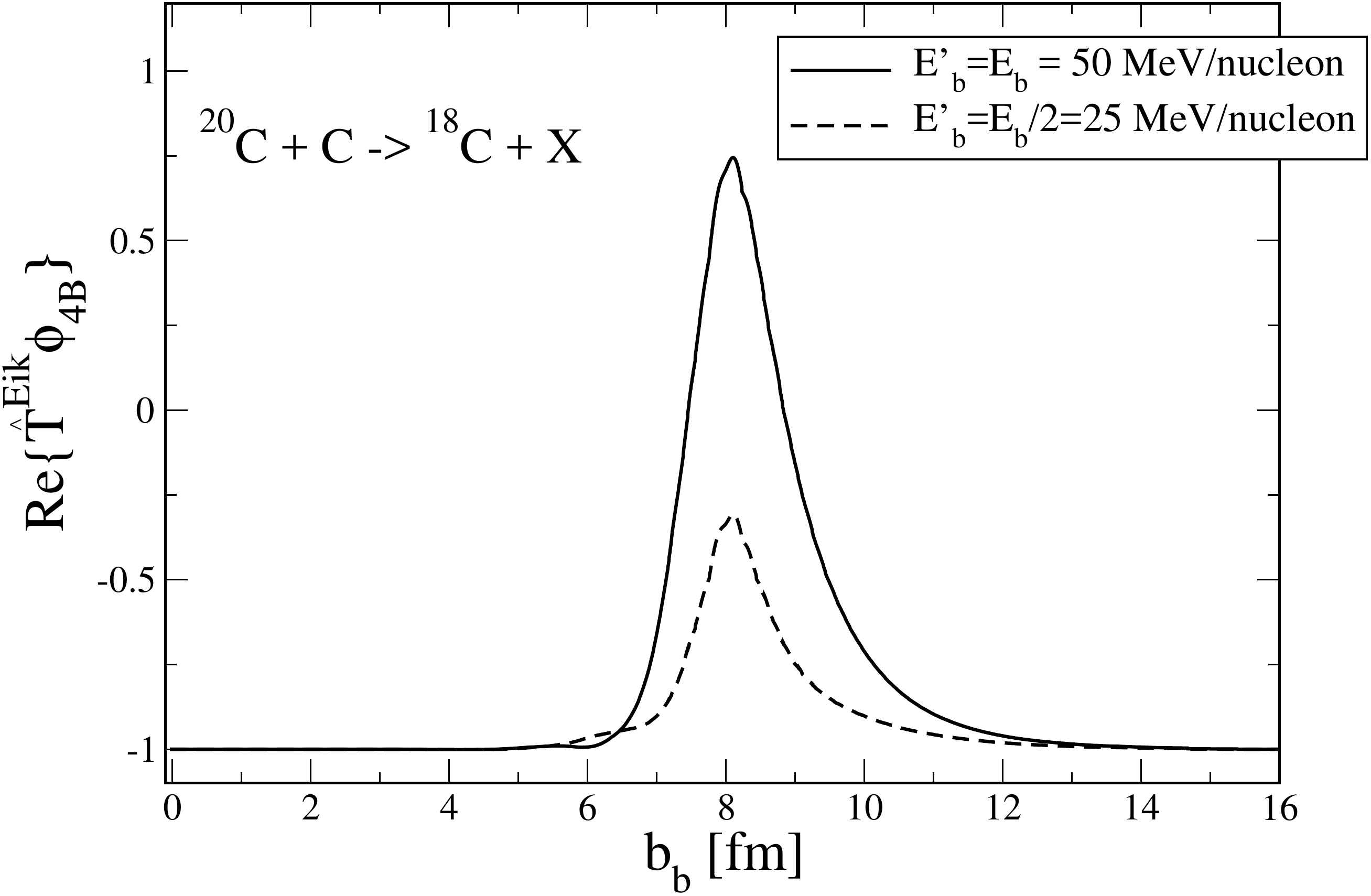}
  \includegraphics[width=0.85\columnwidth]{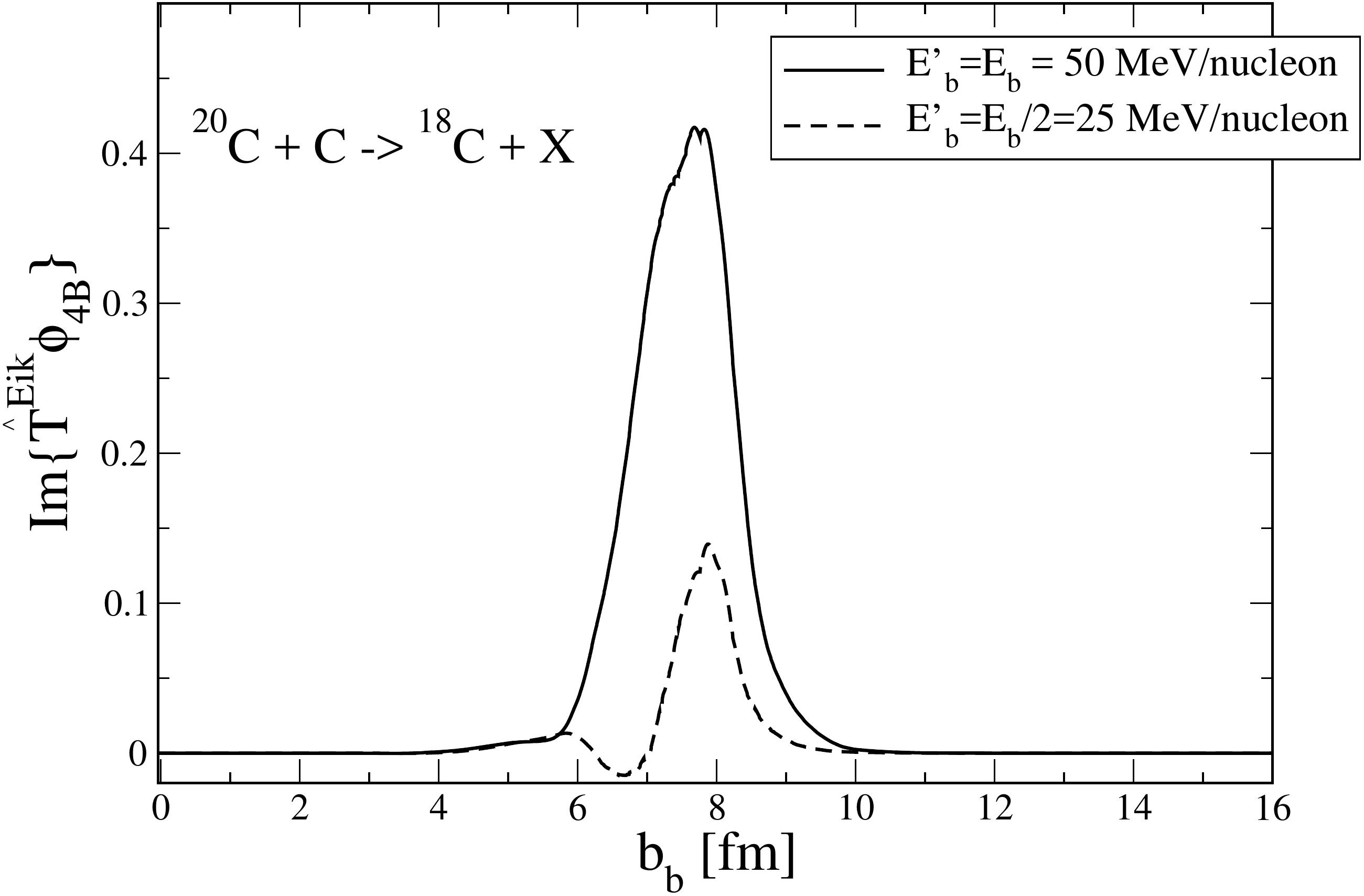}
  \caption{The influence of three-body projectile wave function on $\hat T$-matrix Eq.~(\ref{teik}). Real part is in the  top frame and imaginary at bottom. The neutrons were fixed with $b_1=b_2=8$ fm.}
  \end{figure}

   \begin{figure}[htpb]
  \includegraphics[width=0.85\columnwidth]{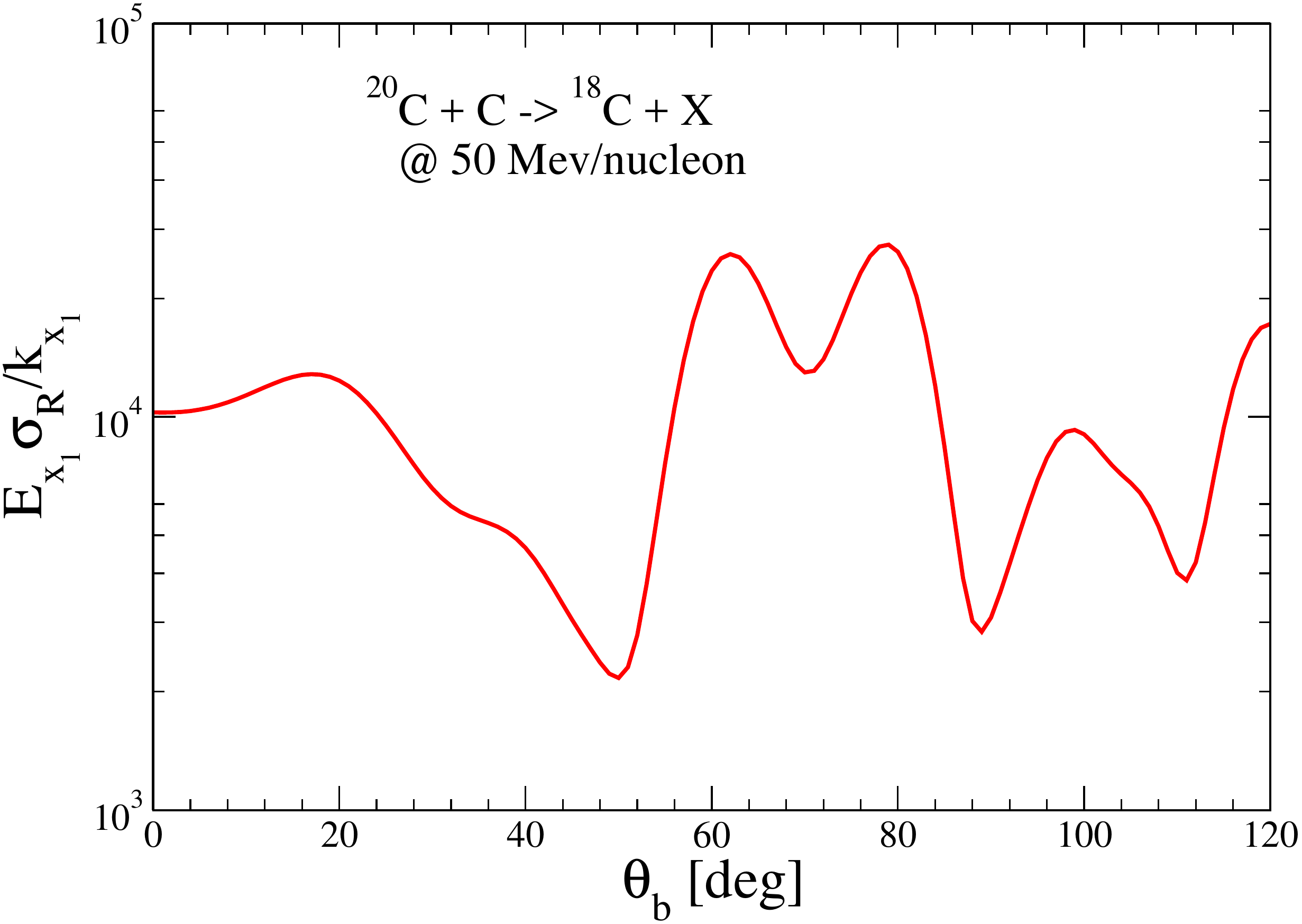}
  \includegraphics[width=0.85\columnwidth]{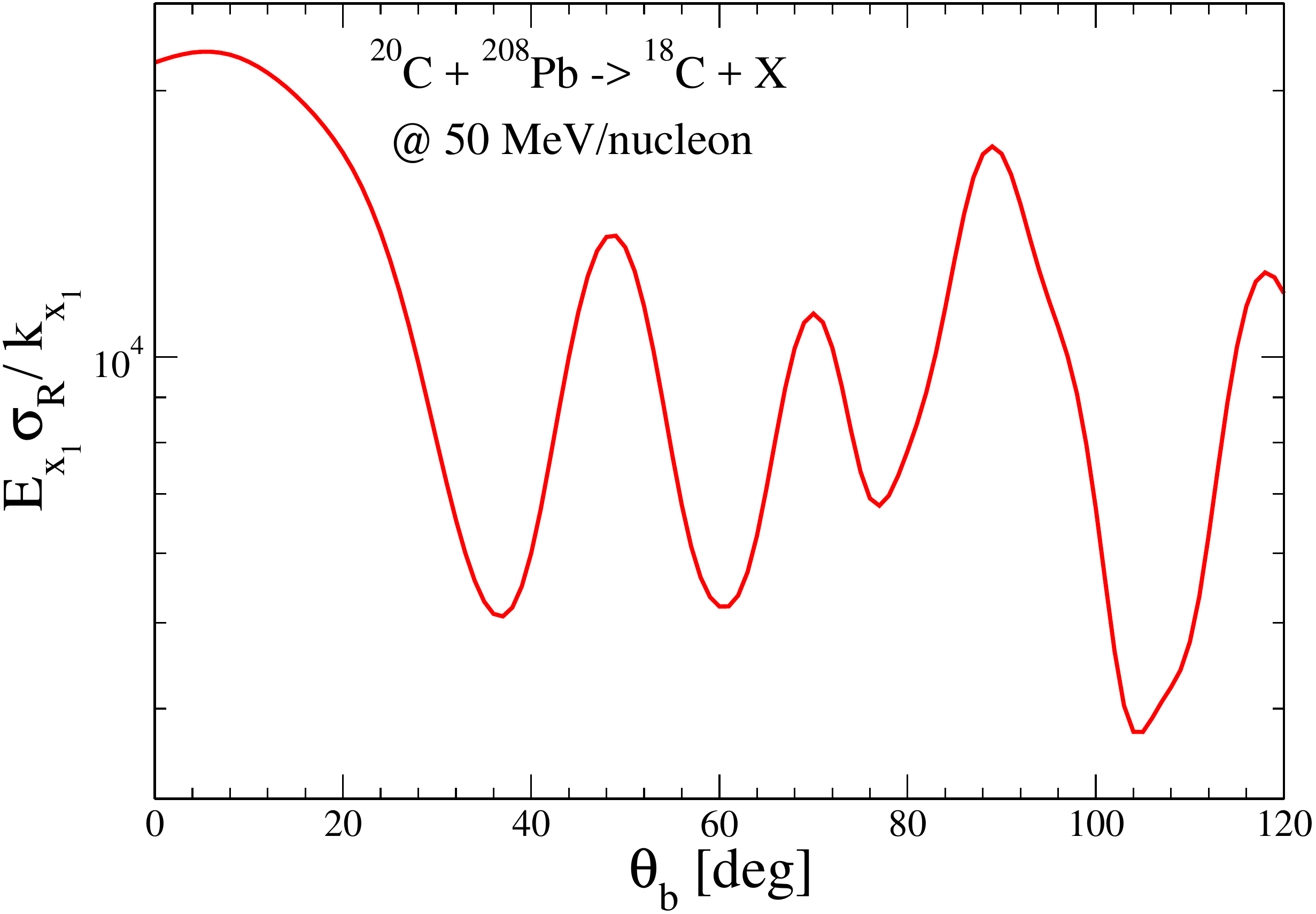}
  \caption{Inclusive four-body fragment cross-section, Eq.(\ref{xx}) as function of scattering angle for fixed core 
  for incoming energy at 50 MeV/nucleon (and outcoming is $E'_b=0.8 E_b$) for targets of $^{12}$C (top) and $^{208}$Pb (bottom).} 
  %Blue line represents the behavior with Coulomb contribution. 
  \label{sigma1}
\end{figure}
The inclusive inelastic cross-section for $x_i$ fragment Eq.~(\ref{xx}) was computed as shown in the Fig.~\ref{sigma1} for targets of $^{208}$Pb and $^{12}$C. 
For both the incoming projectile energy is at 50 MeV/nucleon and $E'_b=0.8 E_b$.
%Coulomb contribution comes from Sao Paulo potential \cite{Chamon} is more evidenced for $^{208}$Pb target while for carbon target, the effect appears to move the cross-section to the right.

\begin{figure}[htpb]
  \includegraphics[width=0.85\columnwidth]{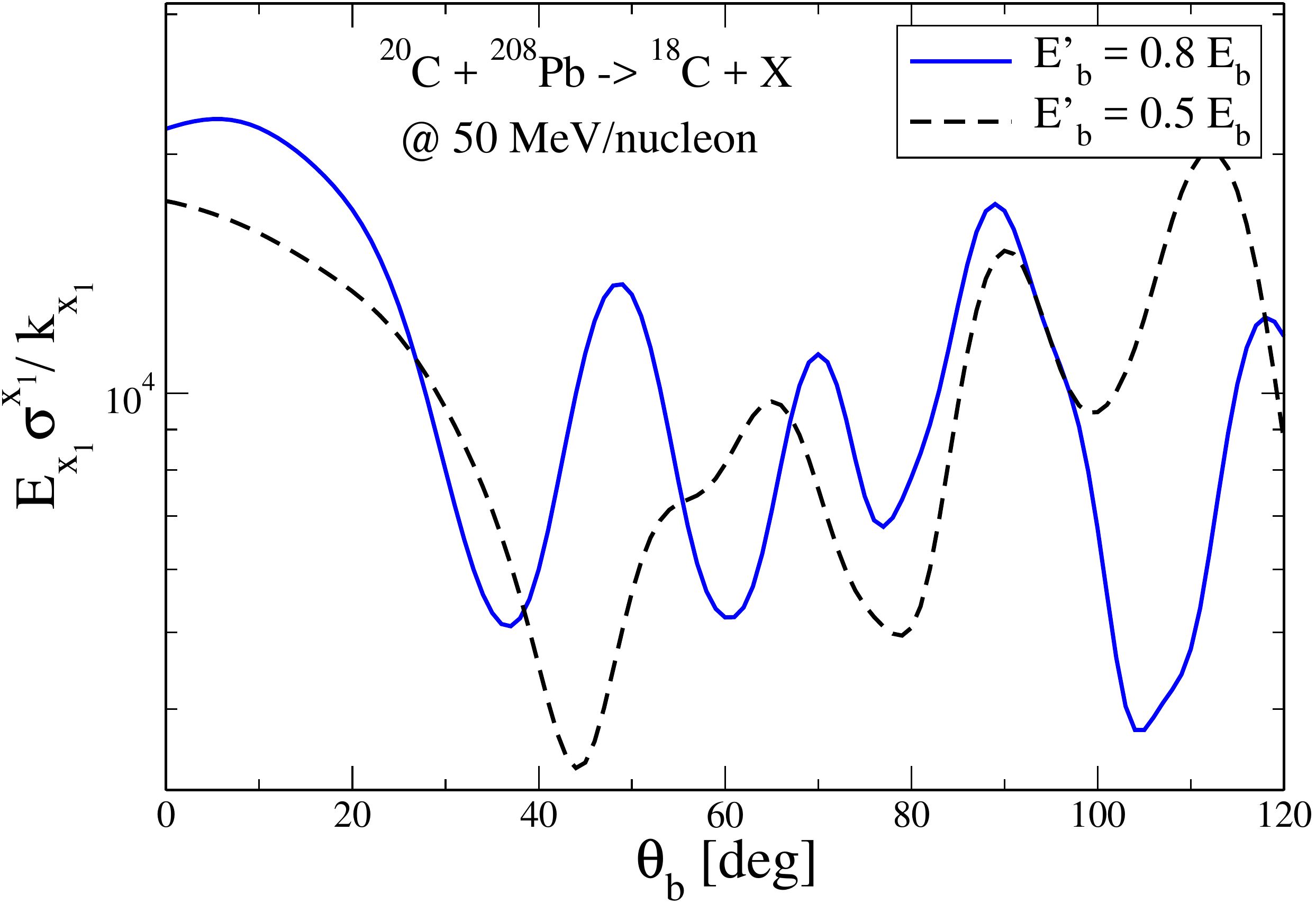}
  \caption{Angular distribution from inclusive four-body fragment cross-section, Eq.(\ref{xx}) for fixed core incoming energy at 50 MeV/nucleon for different outcoming core energies showing the inelastic effect.}
  \label{sigma2}
  \end{figure}

The inelastic effect can be noted by the Fig.~\ref{sigma2}. % which Coulomb scattering is not considered. 
For a fixed incoming energy $E_b=50$ MeV/nucleon, one choose values for outcoming energy such that blue line 
represents $E'_b=0.8 E_b$, and black-dotted line $E'_b=0.5 E_b$. It is possible to see the cross-section oscillation increases as one has a reaction closer to the elastic limit.

  \section{Conclusions}
  The structure of the reaction cross-sections for the absorption of one of the interacting fragments removes the ambiguity about the difference between the 4B and 3B cases, which we find to be damped by the absorption effect of the other fragment. 
  We observed, as shown in the eikonal analysis, the core scattering process is more sensible to the target surface. 
    Although we present preliminary results the structure and results obtained have been helping us to understand the role of the projectile halo in the scattering dynamics. Our next step is the addition the two fragments distorted waves. This will provide a complete description of all ranges of action of the nuclear potential, one may expect it to be very important specially for targets with large number of protons. We are working to add the Coulomb scattering like some works for on neutron-halo \cite{bertulani2004}.
    This interaction plays a role when the three body wave function comes in, small effects can become present specially for large values of impact parameters and forward scattering angles. 
We also plan computing the double fragment inclusive cross-section $\sigma ^{3B}$, providing then, the complete inelastic cross-section.

%  {\it Summary.} sss
  %%%%%% END

  \newpage

  We thank partial support from the Brazilian agencies FAPESP, CNPq and CAPES.   EVC acknowledges financial support from FAPESP grants: 2016/07398-8 and 2017/13693-5.

  \end{document}